\documentclass[
    10pt,
    final,
    journal,
    letterpaper,
    twoside,
    twocolumn,
]{./style/IEEEtran}

\usepackage{epsfig}
\usepackage{subfigure}
\usepackage{amsmath}
\usepackage{amssymb}

\usepackage{psfrag}
\usepackage{cite}
\usepackage{graphicx}
\usepackage{multirow}
\usepackage{array}
\usepackage{enumerate}
\usepackage{graphicx}
\usepackage{times}

\usepackage{balance}   
\usepackage{color}
\usepackage[colorlinks=false, linkcolor=blue, bookmarks=false]{hyperref}

\newtheorem{lemma}{\textbf{Lemma}}

\begin{document}
\title{Joint Link Adaptation and User Scheduling with HARQ in Multi-Cell Environments}
\author{Su~Min~Kim,~\IEEEmembership{Member,~IEEE}, Bang~Chul~Jung,~\IEEEmembership{Senior Member,~IEEE}, and Dan~Keun~Sung,~\IEEEmembership{Fellow,~IEEE}
\thanks{Copyright (c) 2015 IEEE. Personal use of this material is permitted. However, permission to use this material for any other purposes must be obtained from the IEEE by sending a request to pubs-permissions@ieee.org.}
\thanks{Part of this work has been presented at \emph{IEEE International Conference on Communications (ICC)}, Hungary, 2013.}
\thanks{S. M. Kim was with KTH Royal Institute of Technology, Stockholm, Sweden and Ericsson Research, Stockholm, Sweden, and he is currently with the Department of Electronics Engineering, Korea Polytechnic University, Siheung 429-793, Korea (e-mail: \texttt{suminkim@kpu.ac.kr}).}
\thanks{B. C. Jung~(corresponding author) is with the Department of Information and Communication Engineering and Institute of Marine Industry, Gyeongsang National University, Tongyeong 650-160, Korea (e-mail: \texttt{bcjung@gnu.ac.kr}).}
\thanks{D. K. Sung is with the Department of Electrical Engineering, Korea Advanced Institute of Science and Technology (KAIST), Daejeon 305-701, Korea (e-mail: \texttt{dksung@ee.kaist.ac.kr}).}
}
\maketitle

\begin{abstract}
Inter-cell interference~(ICI) is one of the most critical factors affecting performance of cellular networks. In this paper, we investigate a joint link adaptation and user scheduling problem for multi-cell downlink employing HARQ techniques, where the ICI exists among cells. We first propose an approximation method on aggregated ICI for analyzing an effective signal-to-interference-and-noise ratio~(SINR) with the HARQ technique at users, named identical path-loss approximation (IPLA). Based on the proposed IPLA, we propose a transmission rate selection algorithm maximizing an expected throughput at each user. We also propose a simple but effective cross-layer framework jointly combining transmission rate adaptation and user scheduling techniques, considering both HARQ and ICI. 
It is shown that statistical distribution of the effective SINR at users based on the IPLA agrees well with the empirical distribution, while the conventional Gaussian approximation (GA) does not work well in the case that dominant ICIs exist. Thus, IPLA enables base stations to choose more accurate transmission rates. Furthermore, the proposed IPLA-based cross-layer policy outperforms existing policies in terms of both system throughput and user fairness.
\end{abstract}



\section{Introduction}

Compensation for uncertain fading phenomena is one of the most challenging issues in wireless communications. 
In order to improve link reliability and radio resource efficiency against such uncertainty, a \emph{hybrid automatic repeat request (HARQ)} technique has been proposed in physical~(PHY) layer \cite{p_IBM70_Rocher,p_TC85_Chase,p_TC83_Wang}.
Meanwhile, in medium access control (MAC) layer, \emph{dynamic link adaptation} \cite{p_TC98_Goldsmith,p_JSAC99_Balachandran} and \emph{user scheduling} \cite{b_WirelCommun_Goldsmith,b_FundWirelCommun_Tse} techniques have been exploited by using channel state information (CSI) at the transmitter for point-to-point and multi-user environments, respectively. There have existed many studies on link adaptation considering HARQ for various fading channel models in point-to-point communications \cite{p_TWC08_Kim,p_ICC08_Narasimhan,p_TC10_Wu,p_TWC11_Kim,p_ICC11_Kim}. Moreover, several user scheduling algorithms considering HARQ have been proposed for multi-user environments in single-cell networks \cite{p_TWC05_Huang,p_TWC05_Zheng,p_TWC08_Rui,p_ISCIT12_Kim,p_TVT13_Kim}.
Through these studies considering HARQ techniques, the resource efficiency of wireless communications has been improved.

By the way, inter-cell interference (ICI) is another key factor to determine overall system performances in multi-cell networks. In traditional CDMA-based cellular networks \cite{s_UMTS}, the ICI was regarded as an additional source that deteriorates the performance in addition to intra-cell interference among users, which is typically managed by spectrum spreading (i.e., interference averaging) and power control techniques \cite{s_UMTS_SM}. In OFDM-based cellular networks like 3GPP LTE \cite{s_LTE}, however, it has been observed that the ICI significantly degrades the system performance, and thus, many techniques are being proposed in order to mitigate the ICI for OFDM-based cellular networks~\cite{p_CST13_Tafazolli}.  
Especially in heterogeneous network environments, there may exist dominant interferers which significantly affect adjacent cells~\cite{p_WC11_Damnjanovic}.
Therefore, the ICI needs to be carefully managed in the OFDM-based multi-cell networks through efficient link adaptation and user scheduling algorithms.

Recently, there have been several studies on HARQ-based multi-user systems in the presence of interference. Narasimhan analyzed throughput performance of the two-user interference channel with receiver cooperation~\cite{p_ICC10_Narasimhan}. Denic proposed a robust HARQ-incremental redundancy (IR) scheme in the presence of unknown interference such as jamming~\cite{p_ISIT11_Denic}. 
For multiple-input multiple-output (MIMO)-based HARQ systems, several HARQ techniques were proposed by taking into account inter-carrier interference
and inter-antenna interference \cite{p_TVT09_Juang,p_TWC10_Ait-Idir,p_WC07_Park}. 
R\'{a}cz \emph{et al.} investigated an inter-cell interference coordination (ICIC) technique in the uplink 3GPP LTE system, considering HARQ techniques~\cite{p_GC08_Racz}. 
Makki \emph{et al.} \cite{p_CL14_Makki} proposed a coordinated HARQ scheme which reallocates the spectrum of a successfully transmitted user to a user requiring subsequent retransmissions in cooperative multi-cell networks.
Shirani-Mehr \emph{et al.} proposed an optimal scheduling algorithm based on game theory in a multi-user (MU)-MIMO system with HARQ technique in the presence of ICI~\cite{p_TC11_Shirani-Mehr}. They investigated a joint optimization of user scheduling and transmit beamforming with HARQ in a distributed manner. 
To the best of our knowledge, however, there has been no such study that jointly investigates link adaptation and user scheduling with the HARQ technique in a multi-cell environment.

In this paper, we investigate a joint link adaptation and user scheduling problem in a multi-cell downlink network, taking both HARQ and ICI into account.
Main contributions of the paper are summarized as follows:
\begin{itemize}
\item A novel approximation model on the aggregated ICI at each user is proposed for enabling each base station (BS) to determine the optimal transmission rate, which assumes the dominant interfering terms at each user have identical path-loss statistics. Thus, we call it identical path-loss approximation (IPLA) method. The effectiveness of IPLA is examined by comparing it with the conventional approximation method, i.e., Gaussian approximation (GA).
\item An optimal rate selection algorithm with IPLA is proposed for maximizing the expected throughput of a single link in the multi-cell environment. Then, a simple but effective cross-layer framework is also proposed, which jointly combines link adaptation and user scheduling with the HARQ technique for the multi-cell environments. 
\item The performance of the proposed cross-layer framework is evaluated in terms of cell throughput and user fairness through extensive system-level simulations.
\end{itemize}
From the performance evaluation, it is shown that the well-known GA on ICI is not accurate in link adaptation and user scheduling with HARQ for multi-cell environments with some dominant interferers, while the proposed IPLA is highly accurate on the aggregated ICI and thus, it provides an efficient joint link adaptation and user scheduling policy.

The rest of the paper is organized as follows. In Section~\ref{SEC:ICI:System_Model}, the system model is introduced. In Section~\ref{SEC:ICI:LA}, we propose an optimal link adaptation~(transmission rate selection) algorithm for a single link by considering both the HARQ technique and the ICI. In Section~\ref{SEC:ICI:Sch}, we propose a cross-layer framework jointly combining link adaptation and user scheduling and compares the proposed framework with the conventional strategies. In Section~\ref{SEC:ICI:Numerical_Results}, we show the performance of the proposed framework in terms of cell throughput and fairness among users. Finally, we present concluding remarks in Section~\ref{SEC:ICI:Conclusions}.

\section{System Model}\label{SEC:ICI:System_Model}

\begin{figure}
    \centering
    \includegraphics[width=\columnwidth]{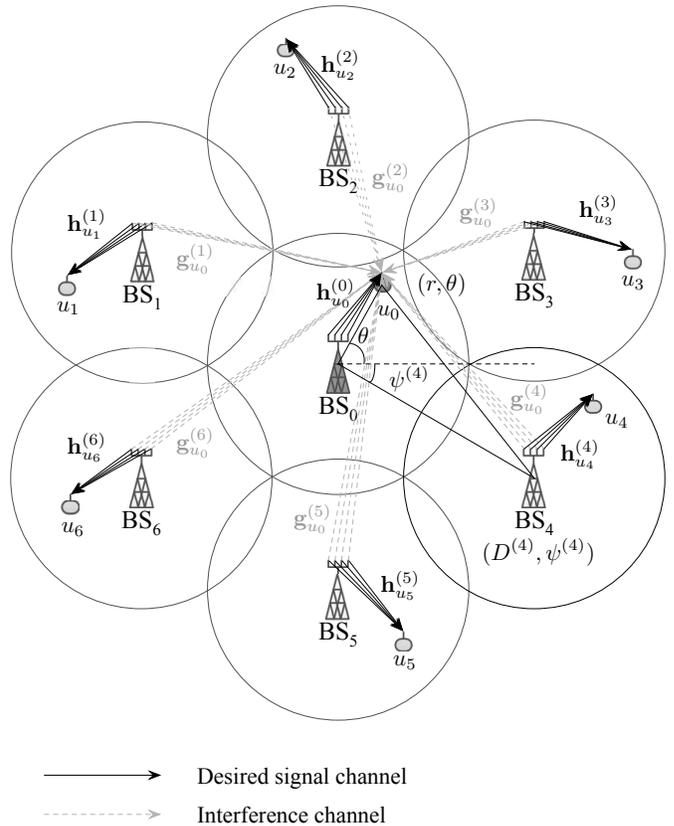}
    \caption{System Model}
    \label{fig:sys_model}
\end{figure}

Fig.~\ref{fig:sys_model} illustrates the system model considered in this paper. We take into account a multi-cell downlink network where there exist $(K+1)$ base stations (BSs) with $M$ transmit antennas and $N$ users with a single receive antenna in each cell. Each BS is assumed to selects a single user for data transmission in this paper for simplicity. In Fig.~\ref{fig:sys_model},  the BS in the center, called home cell, is denoted by superscript $(0)$ and BSs in other cells are denoted by superscript $(k)$, $k \in \{1,\ldots,K\}$. Each BS selects a user within its coverage at each time slot (or scheduling interval), and transmits data with a \emph{random beamforming (RBF)} technique which is also called \emph{opportunistic beamforming (OBF)}
\cite{p_TIT02_Viswanath, p_TIT05_Sharif}. As known in the literature, the RBF technique can achieve the system throughput with \emph{true beamforming}
when sufficiently large number of users exist in a cell, while it can significantly reduce the signalling overhead such as full CSI feedback for the true beamforming. 

The scheduled user in each cell receives a desired signal from its corresponding BS and the ICI signals from $K$ other cell BSs. We focus on the user in the home cell without loss of generality.
The received signal of the scheduled user in the home cell (i.e., $k=0$) is expressed as
\begin{align}
y_{u_0}^{(0)} &= \mathbf{h}_{u_0}^{(0)} \mathbf{v}_{u_0}^{(0)} x_{u_0}^{(0)} + \sum_{k=1}^{K} \mathbf{g}_{u_0}^{(k)} \mathbf{v}_{u_k}^{(k)} x_{u_k}^{(k)} + n_{u_0}^{(0)},
\label{eq:received_signal}
\end{align}
where $u_k$ denotes the index of the selected user in the $k$-th cell. $\mathbf{h}_{u_0}^{(0)}$ and $\mathbf{g}_{u_0}^{(k)}$ denote the channel vectors of the desired signal from the home cell (i.e., $k=0$) and interference signal from the $k$-th cell, respectively. $\mathbf{v}_{u_k}^{(k)}$ indicates the RBF vector for user $u_k$ in the $k$-th cell and $x_{u_k}^{(k)}$ represents the source data symbol of user $u_k$ in the $k$-th cell, and $n_{u_0}^{(0)}$ denotes the additive white Gaussian noise (AWGN), i.e., $n_{u_0}^{(0)} \sim \mathcal{CN}(0,N_0)$ where $N_0$ denotes the noise variance.

In Eq.~(\ref{eq:received_signal}), $\mathbf{h}_{u_0}^{(0)}$ and $\mathbf{g}_{u_0}^{(k)}$ denote the MISO channel vectors including large-scale and small-scale fading components, i.e., 
\begin{align}
\mathbf{h}_{u_0}^{(0)} &\triangleq \left[\sqrt{L_{u_0}^{(0)}}h_{u_0,1}^{(0)}, \sqrt{L_{u_0}^{(0)}}h_{u_0,2}^{(0)}, \cdots, \sqrt{L_{u_0}^{(0)}}h_{u_0,M}^{(0)} \right], \\
\mathbf{g}_{u_0}^{(k)} &\triangleq \left[\sqrt{L_{u_0}^{(k)}}g_{u_0,1}^{(k)}, \sqrt{L_{u_0}^{(k)}}g_{u_0,2}^{(k)}, \cdots, \sqrt{L_{u_0}^{(k)}}g_{u_0,M}^{(k)} \right],
\end{align}
where $h_{u_0,m}^{(0)}$ and $g_{u_0,m}^{(k)}$ denote the small-scale fading signal term of user $u_0$ from the $m$-th antenna of the home BS and the $k$-th BS, respectively, and they are assumed to follow a circular symmetric complex Gaussian distribution with zero mean and unit variance, i.e., $h_{u_0,m}^{(0)}\sim\mathcal{CN}(0,1)$ and $g_{u_0,m}^{(k)}\sim\mathcal{CN}(0,1)$ for $m=\{1,\ldots,M\}$. We assume a slowly varying channel condition and, thus, $h_{u_0,m}^{(0)}$ and $g_{u_0,m}^{(k)}$ are quasi-static during a single HARQ retransmission process. $L_{u_0}^{(0)}$ and $L_{u_0}^{(k)}$ denote the large-scale fading power terms regarded as path-loss of user $u_0$ from the home BS and the $k$-th BS, respectively. 
$L_{u_0}^{(k)}$ is given by $10^{-\frac{\text{PL}_0}{10}}\cdot\left(d_0/d_{u_0}^{(k)}\right)^{\alpha}$, $(k=0,\ldots,K)$
where PL$_0$ denotes the path-loss in dB at reference distance $d_0$, $\alpha$ denotes the path-loss exponent, and $d_{u_0}^{(k)}$ represents the distance between user $u_0$ and the $k$-th BS. When user and BS locations in the home cell are given by $(r,\theta)$ and $(D^{(k)},\psi^{(k)})$, $d_{u_0}^{(k)}$ can be calculated by $\sqrt{r^2 + (D^{(k)})^2 - 2rD^{(k)}\cos(\theta - \psi^{(k)})}$, $(k=0,\ldots,K)$
where $D^{(0)}=0$ (i.e., $d_{u_0}^{(0)}=r$). 
In general, since BSs are deployed at fixed locations in advance, the home BS can easily know the location information of neighboring BSs.  Additionally, we assume that the home BS also knows the user location information through periodic measurement or feedback from the user.
The 3GPP LTE system has been already supporting several user positioning methods even if global positioning system (GPS) signal is unavailable \cite{s_3GPP_36305}. Since we assume the perfect positioning in this paper, the performance can be degraded when the estimated user location is imperfect. The effect of position estimation error is beyond scope of this paper.

Since the RBF scheme is considered as in \cite{p_TIT02_Viswanath}, the beamforming vectors in Eq.~(\ref{eq:received_signal}) are obtained by $\mathbf{v}_{u_k}^{(k)} = \left[ v_{u_k,1}^{(k)}, v_{u_k,2}^{(k)}, \ldots, v_{u_k,M}^{(k)} \right]^T$, $(k=0,\ldots,K)$
where $[\cdot]^T$ denotes the transpose of a vector, $v_{u_k,m}^{(k)} = \sqrt{a_m}e^{j\theta_m}$ where $a_m \in [0, 1]$, $\theta_m \sim \text{Uniform}[-\pi, \pi]$, and $\|\mathbf{v}_{u_k, m}^{(k)}\|^2=\sum_{m=1}^{M}a_m=1$.
Through a property of the RBF scheme, the second term in Eq.~(\ref{eq:received_signal}), the sum of ICI terms, is derived by
\begin{align}\label{eq:ICI}
&\mathcal{I} = \sum_{k=1}^{K} \mathbf{g}_{u_0}^{(k)} \mathbf{v}_{u_k}^{(k)} x_{u_k}^{(k)} \nonumber\\
&= \sum_{k=1}^{K} \sqrt{L_{u_0}^{(k)}} \left( g_{u_0,1}^{(k)} v_{u_k,1}^{(k)} + \cdots + g_{u_0,M}^{(k)}v_{u_k,M}^{(k)} \right) x_{u_k}^{(k)} \nonumber\\
&= \sum_{k=1}^{K} \sqrt{L_{u_0}^{(k)}} \left( \underbrace{ \sqrt{\alpha_1} e^{j\theta_1} g_{u_0,1}^{(k)} + \cdots + \sqrt{\alpha_M} e^{j\theta_M} g_{u_0,M}^{(k)} }_{\sim \mathcal{CN}(0,1)} \right) x_{u_k}^{(k)} \nonumber\\
&= \sum_{k=1}^{K} \sqrt{L_{u_0}^{(k)}} w_{u_0}^{(k)} x_{u_k}^{(k)},
\end{align}
where $w_{u_0}^{(k)}\sim\mathcal{CN}(0,1)$ and the last equality is derived from an isotropic property of complex Gaussian random variable \cite{b_FundWirelCommun_Tse}.

Consequently, the received SINR of the scheduled user in the home cell is
\begin{align}
\gamma = \frac{ s }{\mathcal{X}+ 1/\rho},
\end{align}
where $s=\| \mathbf{h}_{u_0}^{(0)} \mathbf{v}_{u_0}^{(0)} \|^2=L^{(0)}_{u_0}|w^{(0)}_{u_0}|^2$, $\mathcal{X} = \sum_{k=1}^{K}  L_{u_0}^{(k)}  |w_{u_0}^{(k)}|^2$, and $\rho=\frac{P_x}{N_0}$ where $\mathbb{E}[|x^{(k)}_{u_k}|^2]=P_x$ for $k=\{0,\ldots,K\}$. Here, $L_{u_0}^{(0)}$ is a known constant based on user location information at the home BS and we assume that the effective channel power gain of $u_0$, $|w_{u_0}^{(0)}|^2$, is perfectly known at the transmitter (i.e., home BS). It is reasonable because the home BS knows the RBF vector (i.e., $\mathbf{v}_{u_0}^{(k)}$) for user $u_0$ in advance and we consider a quasi-static channel condition where it is possible to estimate the effective channel power gain perfectly for the desired signal channel (i.e., $\mathbf{h}_{u_0}^{(0)}$). Hence, the desired signal power term $s$ is a known constant at the transmitter. Furthermore, the inverse of transmit SNR term is negligible in interference-limited regime (i.e., high SNR regime).

Throughout this paper, we consider the Chase combining based  HARQ (HARQ-CC) protocol, in which every retransmitted information is same as the one at the initial transmission.
The HARQ-CC protocol is simple but obtains a sufficient benefit of HARQ from the combined power gain. Thus, it is widely used in practical wireless communication systems.
\section{Optimal Rate Selection for a Single Link}\label{SEC:ICI:LA}

In this section, we first mathematically formulate the effective SINR and delay-limited throughput (DLT) which represents an expected throughput under a given maximum allowable number of transmissions in HARQ-based systems \cite{p_TWC11_Kim, p_TWC08_Kim, p_ICC08_Narasimhan,p _ISIT08_Narasimhan}.
Then, we consider a well-known Gaussian approximation (GA) on the ICI with noise at users.
Finally, we propose an identical path-loss approximation (IPLA) on the aggregated ICI term at users. 
We also obtain the optimal transmission rate maximizing the DLT for both approximation methods. 

\subsection{Effective SINR and Delay-Limited Throughput}
First of all, the effective SINR after the $n$-th transmission attempt after HARQ-CC\footnote{The HARQ-CC has been widely adopted in 3GPP HSPA \cite{s_HSDPA}, WiMAX \cite{s_WiMAX}, 3GPP LTE \cite{s_LTE}, and their evolutions.} combining becomes
\begin{align}
\gamma(n) = \sum_{i=1}^{n} \gamma_i = \sum_{i=1}^{n}\frac{s}{\mathcal{X}_i + 1/\rho },
\end{align}
where $\gamma_i$ represents the received SINR at the $i$-th transmission, $\mathcal{X}_i$ denotes the ICI power at the $i$-th transmission, which is expressed as $\mathcal{X}_i = \sum_{k=1}^{K} L_{u_0}^{(k)}|w_{u_0}^{(k)}(t_i)|^2$ where $t_i$ indicates the time slot index of the $i$-th transmission, and $s=L^{(0)}_{u_0}|w^{(0)}_{u_0} (t_i)|^2=L^{(0)}_{u_0}|w^{(0)}_{u_0} (t_1)|^2$ denotes the desired signal power. Since the RBF vector is kept during retransmissions and the desired signal channel is quasi-static during retransmissions, $s$ is a known constant for every (re)transmission of a single packet. In contrast, beamforming vectors (i.e., $\mathbf{v}_{u_k}^{(k)}$) in other cells are independently varying according to scheduling decisions by other-cell BSs although interference channel vectors $\mathbf{g}_{u_0}^{(k)}$ are quasi-static during their own retransmission processes.
More specifically, new users can be scheduled after their own transmission successes in other cells during retransmissions in the home cell. This causes asynchronous scheduling  among different cells, which implies that different cells suffer from different user scheduling instances.
Accordingly, the aggregated ICI term is independently varying for every (re)transmission due to the independently varying other-cell beamforming vectors.

Next, the distribution of the effective SINR needs to be analyzed for transmission rate selection. We start to derive the distribution of the effective SINR based on numerical inversion of characteristic function from the following lemma.
\begin{lemma}[Inversion formula of Gil-Pelaez \cite{p_Biom51_Gil-Pelaez}] Let $\phi(t)=\int_{-\infty}^{\infty} e^{jtx} dF(x)$ be a characteristic function (CF) of the one-dimensional distribution function $F(x)$.
For $x$ being the continuity point of the distribution, the following inversion formula holds true:
\begin{align}
F(x)  &= \frac{1}{2} - \frac{1}{\pi} \int_{0}^{\infty} \left( \frac{e^{-jtx} \phi(t) - e^{jtx} \phi(-t)}{2jt} \right) dt \nonumber\\
&= \frac{1}{2} - \frac{1}{\pi} \int_{0}^{\infty} \mathcal{I}m\left( \frac{e^{-jtx} \phi(t)}{t} \right) dt.
\label{eq:cdf}
\end{align}
where $\mathcal{I}m\{\cdot\}$ denotes the imaginary part of a complex number.
\begin{proof}
Refer to \cite{p_Biom51_Gil-Pelaez}.
\end{proof}
\end{lemma}\vspace{0.2in}

By using Lemma~1, if we know the CF of the effective SINR after the $n$-th (re)transmission which is denoted by $\phi_{\gamma(n)}(t)$, we can obtain the cumulative distribution function (CDF), $F_{\gamma(n)}(x)$.
Assuming the information-theoretic capacity-achieving channel coding scheme, the outage probability after the $n$-th (re)transmission is defined by
\begin{align}
P_{out} (n, R) \triangleq \text{Pr} \left\{ \log_2 \left( 1 + \gamma(n) \right) < R \right\} = F_{\gamma(n)} \left( 2^R - 1 \right),
\label{eq:pout}
\end{align}
where $R$ denotes the required transmission source rate.
Then, the DLT is obtained by \cite{p_TWC11_Kim}
\begin{align}
S(R) \triangleq \sum_{i=1}^{N_{max}} \frac{R}{i} \Big[ P_{out}(i-1, R) - P_{out}(i, R) \Big],
\label{eq:dlt}
\end{align}
where $N_{max}$ denotes the maximum allowable number of transmissions in an HARQ retransmission process. 
Substituting Eqns.~\eqref{eq:cdf} and \eqref{eq:pout} for Eq.~(\ref{eq:dlt}), the DLT is finally rewritten by
\begin{align}
S(R) &= \sum_{i=1}^{N_{max}} \frac{R}{i\cdot\pi}\cdot \nonumber\\
&\quad \int_{0}^{\infty} \bigg[ \mathcal{I}m \Big\{ \frac{e^{-jt(2^R-1)}}{t} \cdot\left( \phi_{\gamma(i)}(t) - \phi_{\gamma(i-1)}(t) \right)\Big\} \bigg] dt.
\label{eq:dlt_general}
\end{align}

\subsection{Link Adaptation with Gaussian Approximation (GA)}

Traditionally, the sum of ICI terms is widely approximated as a Gaussian distribution by the well-known central limit theorem (CLT) for even six interference components considering 7-cell structured cellular networks \cite{p_TWC07_Choi,p_JSAC11_Zhu}. Therefore, we investigate a rate selection scheme assuming that the ICI plus noise term follows a Gaussian distribution as a conventional link adaptation scheme.
Through the GA, the effective SINR after HARQ-CC combining can be approximated by
\begin{align}
\gamma(n) = \sum_{i=1}^{n} \,\, \frac{s}{\mathcal{X}_i + 1/\rho } &\approx \sum_{i=1}^{n} \tilde{\gamma}_i = \sum_{i=1}^{n}\frac{s}{|Z_i|^2 }
\end{align}
where $Z_i$ denotes a complex Gaussian random variable with zero mean and variance of the sum of ICI and noise powers, i.e., $Z_i \sim \mathcal{CN}\left(0, \sigma_{Z}^2\right)$ where $\sigma_{Z}^2=\sum_{k=1}^{K} L_{u_0}^{(k)} + 1/\rho$, and $s=L^{(0)}_{u_0}|w^{(0)}_{u_0} (t_1)|^2$. Here, $\tilde{\gamma}_i$ is an inverted Gamma random variable with shape parameter 1 and scale parameter $s/\sigma_{Z}^2$, $\tilde{\gamma}_i$ $\sim$ Inv-Gamma$\left(1, s/\sigma_{Z}^2 \right)$. 

First of all, the PDF of $\tilde{\gamma}_i$ by the GA is
\begin{align}
f_{\tilde{\gamma}_i}(x) = \left(s/\sigma_{Z}^2\right) x^{-2} e^{-s/(\sigma_{Z}^2 x)}.
\end{align}
To derive the distribution of sum of $\tilde{\gamma}_i$ (i.e., $\gamma(n)$), the CF of $\tilde{\gamma}_i$ is derived first as follows: \cite{p_Kyb01_Witkovsky}
\begin{align}
\phi_{\tilde{\gamma}_i}(t) &= \sqrt{-\frac{4jst}{\sigma_{Z}^2}} K_{1}\left( \sqrt{-\frac{4jst}{\sigma_{Z}^2}} \right),
\end{align}
where $K_{\nu}(\cdot)$ denotes the modified Bessel function of the second kind.
Since $\tilde{\gamma}_i$'s are identically and independently distributed (i.i.d.) random variables due to independently varying $\mathcal{X}_i$, the CF of $\gamma(n)$ is obtained by
\begin{align}
\phi_{\gamma(n)}(t) &= \prod_{i=1}^{n} \phi_{\tilde{\gamma}_i}(t) = \left[ \sqrt{-\frac{4jst}{\sigma_{Z}^2}} K_{1}\left( \sqrt{-\frac{4jst}{\sigma_{Z}^2}} \right) \right]^n.
\label{eq:cf_ga}
\end{align}
By substituting Eq.~(\ref{eq:cf_ga}) for Eq.~(\ref{eq:dlt_general}), the DLT of the conventional GA is obtained by
\begin{align}
S_{\textrm{GA}}(R) &= \sum_{i=1}^{N_{max}} \frac{R}{i\cdot\pi} \cdot \nonumber\\
&\quad \int_{0}^{\infty} \bigg[ \mathcal{I}m \Big\{ \Psi_{\textrm{GA}}(t;i, R)  - \Psi_{\textrm{GA}}(t;i-1,R) \Big\} \bigg] dt,
\label{eq:dlt_ga}
\end{align}
where
\begin{align}
\Psi_{\textrm{GA}}(t;i,R) =  \frac{ e^{-jt(2^R-1)} }{ t } \left[ \sqrt{ - \frac{ 4jst }{ \sigma_{Z}^2 } } K_{1} \left( \sqrt{ - \frac{ 4jst }{ \sigma_{Z}^2 } } \right) \right]^k. \nonumber
\end{align}
Finally, the optimal source rate for maximizing the DLT through the conventional GA is determined by
\begin{align}
R_{\textrm{GA}}^* &= { \mathrm{argmax}\atop{\scriptstyle{R \geq 0}} }~ \sum_{i=1}^{N_{max}} \frac{R}{i\cdot\pi} \cdot \nonumber\\
&\quad \int_{0}^{\infty} \bigg[ \mathcal{I}m \Big\{ \Psi_{GA}(t;i, R) - \Psi_{GA}(t;i-1,R) \Big\} \bigg] dt.
\label{eq:ra_ga}
\end{align}

\subsection{Link Adaptation with Identical Path-Loss Approximation (IPLA)}

Assuming the interference-limited regime (i.e., $\rho \gg 1$), the effective SINR after HARQ-CC combining can be approximated by
\begin{align}
\gamma(n) &\approx \sum_{i=1}^{n}\frac{s}{\mathcal{X}_i } = \sum_{i=1}^{n}\frac{s}{\sum_{k=1}^{K}  L_{u_0}^{(k)}  |w_{u_0}^{(k)} (t_i) |^2},
\label{eq:sir}
\end{align}
where $t_i$ denotes the time slot index of the $i$-th transmission and $s=L^{(0)}_{u_0}|w^{(0)}_{u_0} (t_1)|^2$. The sum of ICI terms in the denominator of Eq.~\eqref{eq:sir} is a weighted sum of Gamma random variables since $w_{u_0}^{(k)} (t_i)\sim\mathcal{CN}(0,1)$ and $L_{u_0}^{(k)} \neq L_{u_0}^{(l)}$ for $k\neq l \in \{1,\ldots,K\}$. 
Note that there exist no closed-form expression for such distribution even though there have been some efforts to develop computational methods \cite{Mat82AISM,MC84CMA}.
Furthermore, the distribution of the effective SINR after the $n$-th transmission attempt, $\gamma(n)$, which is the sum of inverse of the weighted sum of Gamma random variables, has a much more complicated form and therefore it is intractable to derive its CF mathematically. Since only the contribution of the aggregated ICI rather than individual ICIs is interested in the effective SINR and even non-dominant interferers cannot be simply negligible\footnote{To validate this statement, we examine the expected throughput of taking three dominant interferers among whole interferers, compared with the exact one in Section~\ref{SEC:link_throughput}}, we propose to approximate all path-loss terms from other-cell BSs to be identical as their average value. Then, the effective SINR can be approximated by
\begin{align}
\gamma(n) \approx \sum_{i=1}^{n}\tilde{\gamma}_i= \sum_{i=1}^{n}\frac{s}{ \sum_{k=1}^{K} \bar{L}  |w_{u_0}^{(k)} (t_i) |^2},
\end{align}
where $\bar{L}=\frac{1}{K}\sum_{k=1}^{K}L_{u_0}^{(k)}$ denotes the average value of all path-loss terms from other cells. 
It is worth noting that the proposed IPLA preserves the average statistics of the aggregated ICI since for a given user, $\mathbb{E}\big[\sum_{k=1}^{K} \bar{L}  |w_{u_0}^{(k)} (t_i) |^2\big]=\bar{L}\mathbb{E}\big[\sum_{k=1}^{K}  |w_{u_0}^{(k)} (t_i) |^2\big]=\bar{L}\sum_{k=1}^{K}\mathbb{E}\big[ |w_{u_0}^{(k)} (t_i) |^2\big]=\bar{L}K=\sum_{k=1}^{K}L_{u_0}^{(k)}$, while $\mathbb{E}\big[\sum_{k=1}^{K}  L_{u_0}^{(k)}  |w_{u_0}^{(k)} (t_i) |^2\big]=\sum_{k=1}^{K}  L_{u_0}^{(k)} \mathbb{E}\big[ |w_{u_0}^{(k)} (t_i) |^2\big]=\sum_{k=1}^{K}L_{u_0}^{(k)}$ because $L_{u_0}^{(k)}$'s are deterministic for the given user.

Now, the sum of ICI terms becomes the sum of i.i.d. Gamma random variables and it also follows a Gamma distribution. After all, the approximated SINR at the $i$-th transmission, $\tilde{\gamma}_i$, follows an inverted Gamma distribution, i.e., $\tilde{\gamma}_i$ $\sim$ Inv-Gamma$(K,s/\bar{L})$. 
Hence, the probability density function (PDF) of $\tilde{\gamma}_i$ is
\begin{align}
f_{\tilde{\gamma}_i}(x) = \frac{\left(s/\bar{L}\right)^{K}}{(K-1)!} x^{-K-1} e^{-s/(\bar{L}x)}.
\end{align}
Then, the CF of $\tilde{\gamma}_i$ and $\gamma(n)$ are derived, respectively, as follows: \cite{p_Kyb01_Witkovsky}
\begin{align}
\phi_{\tilde{\gamma}_i}(t) &= \frac{2(-\frac{jst}{\bar{L}})^{\frac{K}{2}}}{(K-1)!} K_{K}\left( \sqrt{-\frac{4jst}{\bar{L}}} \right),\\
\phi_{\gamma(n)}(t) &= \prod_{i=1}^{n} \phi_{\tilde{\gamma}_i}(t) 
= \left[ \frac{2(-\frac{jst}{\bar{L}})^{\frac{K}{2}}}{(K-1)!} K_{K}\left( \sqrt{-\frac{4jst}{\bar{L}}} \right) \right]^n,
\label{eq:cf_ipla}
\end{align}
where $K_{\nu}(\cdot)$ denotes the modified Bessel function of the second kind.
By substituting Eq.~(\ref{eq:cf_ipla}) for  Eq.~(\ref{eq:dlt_general}), the DLT by the proposed IPLA is
\begin{align}
S_{\textrm{IPLA}}(R) &= \sum_{i=1}^{N_{max}} \frac{R}{i\cdot\pi} \cdot\nonumber\\
&  \int_{0}^{\infty} \bigg[ \mathcal{I}m \Big\{ \Psi_{\textrm{IPLA}}(t;i, R) - \Psi_{\textrm{IPLA}}(t;i-1,R) \Big\} \bigg] dt,
\label{eq:dlt_ipla}
\end{align}
where
\begin{align}
\Psi_{\textrm{IPLA}}(t;i,R) = \frac{ e^{-jt(2^R-1)} }{ t } \left[ \frac{ 2\left( - \frac{ jst }{ \bar{L} } \right)^{ \frac{ K }{ 2 } } }{ (K-1)! } K_{K} \left( \sqrt{ - \frac{ 4jst }{ \bar{L} } } \right) \right]^i.\nonumber
\end{align}
Eventually, the optimal source rate based on the proposed IPLA for maximizing the DLT is determined by
\begin{align}
R_{\textrm{IPLA}}^* &= { \mathrm{argmax}\atop{\scriptstyle{R \geq 0}} }~ \sum_{i=1}^{N_{max}} \frac{R}{i\cdot\pi} \cdot\nonumber\\
&\quad \int_{0}^{\infty} \bigg[ \mathcal{I}m \Big\{ \Psi_{\textrm{IPLA}}(t;i, R)  - \Psi_{\textrm{IPLA}}(t;i-1,R) \Big\} \bigg] dt.
\label{eq:ra_ipla}
\end{align}
The above optimal source rate can be easily found by a grid search or a Golden section search with range between zero and a proper upper limit, since the DLT has a shape of quasi-concave function with respect to the source rate as shown numerically in Section~\ref{SEC:link_throughput}, although it cannot be analytically proved due to a sophisticated form of the DLT formula.

\section{Joint Link Adaptatin and User Scheduling: Cross-Layer Framework}\label{SEC:ICI:Sch}

\subsection{Overall Procedure}

We first propose a simple cross-layer framework to perform both link adaptation and user scheduling considering HARQ and ICI.
The cross-layer framework consists of three main components: \emph{rate selection}, \emph{effective rate mapping}, and \emph{scheduler}. The roles of components are described as in the following.

\subsubsection{Rate Selection (RS)}
The RS plays a role to determine an optimal transmission source rate $R_u^*(t)$ for the $u$-th user at initial transmission instance of the HARQ-based system. In this paper, we consider RS schemes to maximize the DLT of each user considering HARQ retransmission and ICI statistics as presented in the previous section.

\subsubsection{Effective Rate Mapping (ERM)}
The transmission source rate is different from the achievable rate in HARQ-based systems due to uncertain retransmissions. Therefore, an effective rate, which is close to the achievable rate, needs to be taken into account for user scheduling if it is available.
The ERM determines an effective rate $R_{\text{eff},u}(t)$ for the $u$-th user as a function of the optimal source rate $R_u^*(t)$, i.e., $R_{\text{eff},u}(t)=f(R_u^*(t))$, in order to adjust the scheduling priority of each user. Through such ERM, the instantaneous rate $R_u(t)$ in the scheduler is replaced by the effective rate $R_{\text{eff},u}(t)$. 
After all, the scheduler selects a user with the highest utility value substituted into the effective rate.

\subsubsection{Scheduler}
The scheduler determines which user is the best at every scheduling instance. There are three representative scheduling algorithms: \emph{Round Robin (RR)}, \emph{Max C/I}, and \emph{Proportional Fair (PF)}.
In this paper, we take into account the PF scheduler for an asymmetric user distribution scenario where users have different distances from the BS, in order to consider user fairness.
The PF scheduler is simply expressed as:
\begin{align}
u^* =  \underset{u\in\Pi}{\mathrm{argmax}}~\frac{R_u(t)}{T_u(t)},
\end{align}
where $\Pi$ denotes the set of users in a cell, $R_u(t)$ denotes the achievable rate of the $u$-th user at time slot $t$, and $T_u(t)$ denotes the average throughput of the $u$-th user at time slot $t$, which is updated as $T_u(t+1) = (1-\frac{1}{t_c})\cdot T_u(t) + \frac{1}{t_c} \cdot R_u(t) \cdot \mathbb{I}\{u=u^*\}$
where $t_c$ denotes the pre-determined windowing interval for moving averaging and $\mathbb{I}\{x\}$ denotes the indication function which is one if $x$ is true and zero otherwise.
In the HARQ-based systems, $R_u(t)$ should be modified considering the HARQ retransmission process and it can be done by the ERM in this cross-layer framework.

The operating procedure according to the proposed cross-layer framework is illustrated as follows:\\\\
\begin{tabular}{|l|}
  \hline
  (Step~1) $[$\textbf{Rate Selection}$]$: Determine $R_u^*(t)$\\
  (Step~2) $[$\textbf{Effective Rate Mapping}$]$: \\
  \quad\quad\quad\quad Determine $R_{\text{eff},u}(t)=f\left(R_u^*(t)\right)$\\
  (Step~3) $[$\textbf{Scheduler}$]$: Determine $u^* = \underset{u\in\Pi}{\mathrm{argmax}}~\frac{R_{\text{eff},u}(t)}{T_u(t)}$\\
  (Step~4) $[$\textbf{HARQ Transmission}$]$\\
      \quad\, -- $u^*$ transmits with $R_u^*(t)$ until successful transmission \\
      \quad\,\quad or maximum transmission limit.\\
      \quad\, -- Go to (Step~1) for all users after the end of the \\
      \quad\,\quad (re)transmissions of the scheduled user $u^*$.\\
  \hline
\end{tabular}

\subsection{Proposed Cross-Layer Policy and Other Candidates}

In this subsection, we propose an IPLA-based cross-layer policy. We also introduce two reference and two conventional policies for the performance comparison in next section. Hereafter, each cross-layer policy is denoted by the rate selection and effective rate mapping, i.e., $\mathcal{P}$\{RS, ERM\}, since all policies employ the same PF scheduler.

\subsubsection{Proposed IPLA-based Policy, $\mathcal{P}$\{$\mathrm{RS}$-$\mathrm{IPLA}$, $S_{\mathrm{IPLA}}(R_{\mathrm{IPLA}}^*)$\}}

        The proposed IPLA-based policy is based on RS through the IPLA on the aggregated ICI term (so called RS-IPLA).
        According to the RS-IPLA proposed in the previous section, the transmission source rate of the $u$-th user is determined by Eq.~(\ref{eq:ra_ipla}).
        Since it takes into account an HARQ retransmission process using statistics of the aggregated ICI term, which is assumed by a Gamma distribution through the IPLA, the achievable rate is different from the transmission source rate. Thus, the \emph{expected throughput}-based ERM is considered for user scheduling as
        \begin{align}
        R_{\text{eff},u} =& S_{\mathrm{IPLA}}(R_{\mathrm{IPLA},u}^*) \nonumber\\
        =& \sum_{i=1}^{N_{max}} \frac{R_{\mathrm{IPLA},u}^*}{i\cdot\pi} \int_{0}^{\infty} \bigg[ \mathcal{I}m \Big\{ \Phi_{\mathrm{IPLA}}(t;i, R_{\mathrm{IPLA},u}^*) \nonumber\\
        &\quad\quad\quad\quad - \Phi_{\mathrm{IPLA}}(t;i-1,R_{\mathrm{IPLA},u}^*) \Big\} \bigg] dt,
        \label{eq:ranking_ipla}
        \end{align}
        where
        \begin{align}
        &\Phi_{\mathrm{IPLA}}(t;i,R_{\mathrm{IPLA},u}^*) \nonumber\\
        &= \frac{ e^{-jt(2^{R_{\mathrm{IPLA},u}^*}-1)} }{ t } \cdot \left[ \frac{ 2\left( - \frac{ jst }{ \bar{L} } \right)^{ \frac{ K }{ 2 } } }{ (K-1)! } \cdot K_{K} \left( \sqrt{ - \frac{ 4jst }{ \bar{L} } } \right) \right]^i.\nonumber
        \end{align}
        Based on the selected transmission source rate and effective rate, the user scheduling is performed with the PF criterion and then, the HARQ transmission is performed for the scheduled user according to (Step~4) in the operating procedure of the cross-layer framework.

\subsubsection{Reference Policies}

\begin{itemize}
    \item \emph{Genie-Aided Policy}, $\mathcal{P}$\{RS-Opt, $R_{\mathrm{Opt}}^*\}$\\
        The genie-aided policy has perfect knowledge on instantaneous ICI terms. In this case, the transmitter can accurately adapt to instantaneous interference channel and the channel capacity depending on the instantaneous interference channel conditions is achieved without any retransmission and outage. Even though this policy is rather unrealistic, it offers an upper bound of the system performance.
        According to the RS-Opt scheme with the perfect knowledge for interference channels, the transmission source rate of the $u$-th user is determined by
        \begin{equation}
        R_{\text{Opt},u}^* = \log_2\left( 1 + \frac{s}{\sum_{k=1}^{K}L_{u}^{(k)} |w_{u}^{(k)}(t_1)|^2 + 1/\rho} \right),
        \end{equation}
        where $t_1$ denotes the time index at initial transmission, $|w_{u}^{(k)}(t_1)|^2$ represents the exact effective interference channel power gain from the $k$-th BS to the $u$-th user in the home cell, $s=L^{(0)}_{u}|w^{(0)}_{u} (t_1)|^2$, and $\rho=\frac{P_x}{N_0}$.
        Next, since the genie-aided policy does not cause outage and retransmission, the \emph{instantaneous rate}-based ERM is considered as
        \begin{equation}\label{eq:ERM_Opt}
        R_{\text{eff},u} = R_{\text{Opt},u}^*.
        \end{equation}
    \item \emph{Instantaneous SINR Policy}, $\mathcal{P}$\{RS-$i$-SINR, $R_{i\text{-}\mathrm{SINR}}^*$\}\\
        The instantaneous SINR policy is the simplest one based on the SINR value fed back from the receiver. This policy has inaccurate rate selection due to independently varying interference channels for every (re)transmission and a feedback delay.
        According to the RS-$i$-SINR scheme using the outdated feedback channel information, the transmission source rate of the $u$-th user is determined by
        \begin{equation}
        R_{i\text{-SINR},u}^* = \log_2\left( 1 + \frac{s}{\sum_{k=1}^{K}L_{u}^{(k)} |w_{u}^{(k)}(t_1 - \delta)|^2 + 1/\rho} \right),
        \end{equation}
        where $|w_{u}^{(k)}(t_1 - \delta)|^2$ represents the interference power gain from the $k$-th BS to user $u$ with the feedback delay $\delta$, $s=L^{(0)}_{u}|w^{(0)}_{u} (t_1)|^2$, and $\rho=\frac{P_x}{N_0}$.
        Since the instantaneous SINR policy takes advantage of only instantaneous information without consideration of HARQ retransmission, the \emph{instantaneous rate}-based ERM is also considered as
        \begin{equation}\label{eq:ERM_i-SINR}
        R_{\text{eff},u} = R_{i\text{-SINR},u}^*.
        \end{equation}
\end{itemize}

\subsubsection{Conventional Policies}

\begin{itemize}
    \item \emph{Average Interference Policy}, $\mathcal{P}$\{RS-Avg-$\mathcal{X}$, $R_{\text{Avg-}\mathcal{X}}^*$\}\\
        The average interference policy exploits an average value for the aggregated ICI term \cite{p_TWC10_Ait-Idir},
        since each ICI term is an uncertain and independently varying factor.
        According to the RS-Avg-$\mathcal{X}$ scheme replacing the aggregated ICI term by the average value, the transmission source rate of the $u$-th user is determined by
        \begin{equation}
        R_{\text{Avg-}\mathcal{X},u}^* = \log_2\left( 1 + \frac{s}{\bar{\mathcal{X}} + 1/\rho} \right),
        \end{equation}
        where $\bar{\mathcal{X}}=\mathbb{E}\left[\sum_{k=1}^{K} L_{u}^{(k)} |w_{u}^{(k)}|^2\right]=\sum_{k=1}^{K} L_{u}^{(k)}$, $s=L^{(0)}_{u}|w^{(0)}_{u}(t_1)|^2$, and $\rho=\frac{P_x}{N_0}$.
        The average interference policy has an identical source rate during retransmissions because of applying the average value for the sum of ICI terms. Thus, similarly to Eqns. \eqref{eq:ERM_Opt} and \eqref{eq:ERM_i-SINR}, the \emph{instantaneous rate}-based ERM is considered as
        \begin{equation}
        R_{\text{eff},u} = R_{\text{Avg-}\mathcal{X},u}^*.
        \end{equation}
    \item \emph{GA-based Policy}, $\mathcal{P}$\{RS-GA, $S_{\mathrm{GA}}(R_{\mathrm{GA}}^*)$\}\\
        The GA-based policy is based on RS through GA for the sum of ICI and noise terms.
        According to the RS-GA scheme investigated in the previous section, the transmission source rate of the $u$-th user is determined by Eq.~(\ref{eq:ra_ga}).
        Since the GA-based policy considers an HARQ retransmission process using statistics of the aggregated ICI term, the \emph{expected throughput}-based ERM is considered as
        \begin{align}
        &\!\!\!\!\!\! R_{\text{eff},u} = S_{\mathrm{GA}}(R_{\mathrm{GA},u}^*)= \sum_{i=1}^{N_{max}} \frac{R_{\mathrm{GA},u}^*}{i\cdot\pi} \cdot  \nonumber\\
        &\!\!\!\!\!\! \int_{0}^{\infty} \bigg[ \mathcal{I}m \Big\{ \Phi_{\mathrm{GA}}(t;i, R_{\mathrm{GA},u}^*) - \Phi_{\mathrm{GA}}(t;i-1,R_{\mathrm{GA},u}^*) \Big\} \bigg] dt,
        \end{align}
        where
        \begin{align}
        &\Phi_{\mathrm{GA}}(t;i,R_{\mathrm{GA},u}^*) \nonumber\\
        &=  \frac{ e^{-jt(2^{R_{\mathrm{GA},u}^*}-1)} }{ t } \cdot \left[ \sqrt{ - \frac{ 4jst }{ \sigma_{Z}^2 } } \cdot K_{1} \left( \sqrt{ - \frac{ 4jst }{ \sigma_{Z}^2 } } \right) \right]^i.\nonumber
        \end{align}
\end{itemize}

\section{Numerical Results} \label{SEC:ICI:Numerical_Results}

In this section, we first examine the effectiveness of the proposed IPLA through a quantile versus quantile (Q-Q) plot on the effective SINR distribution, $F_{\gamma(n)}(x)$.
After that, we discuss effects of user distance and path-loss exponent on the rate selection in a single link.
Finally, we evaluate the performance of the proposed, conventional, and reference cross-layer policies in terms of system throughput and fairness metric, through system-level simulations. As basic simulation setups, we consider an 1-tier cellular network with six other cells (i.e. $K=6$) where users are asymmetrically distributed in the home cell. The BS-to-BS distance is set to $1000$ m (i.e., $D^{(k)}=1000, \; \forall k$) and angles between BS in the home cell and BSs in the other cells are set to $\mathbf{\psi}^{(k)}=\frac{5\pi}{6}-\frac{k\pi}{3}$, i.e., $\vec{\psi}=[\frac{5\pi}{6}, \frac{\pi}{2}, \frac{\pi}{6}, -\frac{\pi}{6}, -\frac{\pi}{2}, -\frac{5\pi}{6}]$. We set the distances between users and the home BS to $r \in [150, 200, 250, 300, 400]$m and each element in the vector is equally set according to the number of users. Therefore, we just consider the number of users as multiple of five and it is 250 m for a single user case. Additionally, the angles between users and the home BS are uniformly determined as $\theta=$Uniform$[-\pi,\pi]$. For path-loss, we set PL$_0$ to 37 dB at reference distance $d_0=1000$ m and path-loss exponent, $\alpha$, to 3. The maximum allowable number of transmissions, $N_{max}$, is set to 4, which is a typical value in LTE and WiMAX systems. In order to take into account an interference-limited situation, we set transmit SNR, $\rho$, to 43 dB.\footnote{In this setting, the average received SNR without interference becomes 6 dB when the distance is 1000 m.}

\subsection{Statistical Distribution of Effective SINR based on IPLA}
In order to examine the effectiveness of the proposed IPLA, we introduce a Q-Q plot which is widely used for quantitative comparison between two distributions.
It can provide an intuitive comparison between two statistical data sets as well as two theoretical distributions and more information on the local agreement between two distributions than other fitting tests such as Chi-square and Kolmogorov-Smirnov tests \cite{b_QQplot_Gibbons}. In this paper, we compare two theoretically approximated distributions with the real empirical distribution. In the Q-Q plot, the $x$-axis is based on the theoretical distribution with the approximated CDF, which is obtained by inverting the CDF, $F_{\gamma(n)}^{-1}(x)$, and the $y$-axis is based on the empirical quantile from a sample data set on the effective SINR obtained by statistical realizations.

\begin{figure}[pt]
    \centering
    \includegraphics[width=\columnwidth]{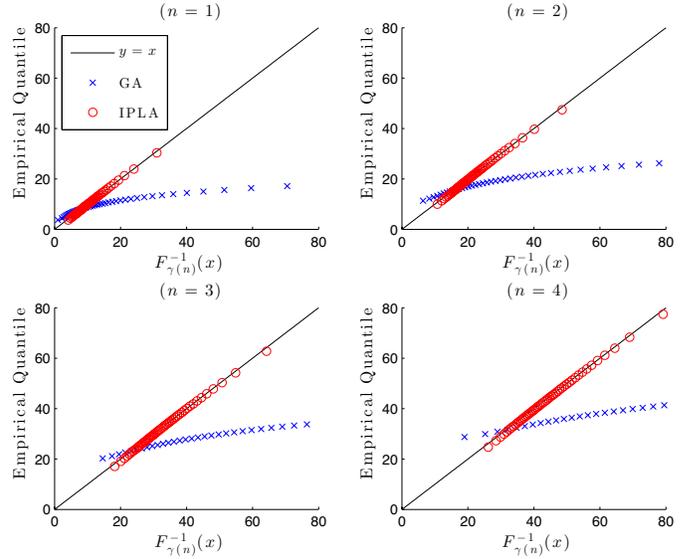}
    \caption{Q-Q plots of the proposed IPLA and the conventional GA ($K=6$, $r=250$ m, $\theta=\pi/2$, $\alpha=3$, $|w_{u_0}^{(0)}|^2=1$, $N_{max}=4$) }
    \label{fig:QQ_plot}
\end{figure}

Fig.~\ref{fig:QQ_plot} shows the Q-Q plots of the proposed IPLA and the conventional GA compared with the real empirical distribution according to the number of transmission attempts of a single packet, $n$. Since the line $y=x$ represents the identity of two compared distributions in the Q-Q plot, the proposed IPLA almost following the $y=x$ line agrees well with the real empirical distribution, regardless of $n$ values. As a general trend, with increasing the number of transmission attempts, $n$, the Q-Q plots of both approximated distributions move to the right-upper side, which implies a larger effective SINR value due to an HARQ-CC combining gain. While the Q-Q plot of the proposed IPLA agrees well, that of the conventional GA is flatter than the line $y=x$. This implies that the approximated distribution by the GA is more dispersed than the real empirical distribution. Additionally, difference between the approximated distribution by the GA and the empirical distribution increases as the value of effective SINR increases.
Through the comparison of the approximated distributions with the real empirical distribution, it is shown that the proposed IPLA offers a good approximation on the effective SINR, while the conventional GA gives significant differences in the approximation.

\subsection{Link Adaptation for a Single Link: Effects of User Distance and Path-Loss Exponent}\label{SEC:link_throughput}

\begin{figure}
    \centering
    \includegraphics[width=\columnwidth]{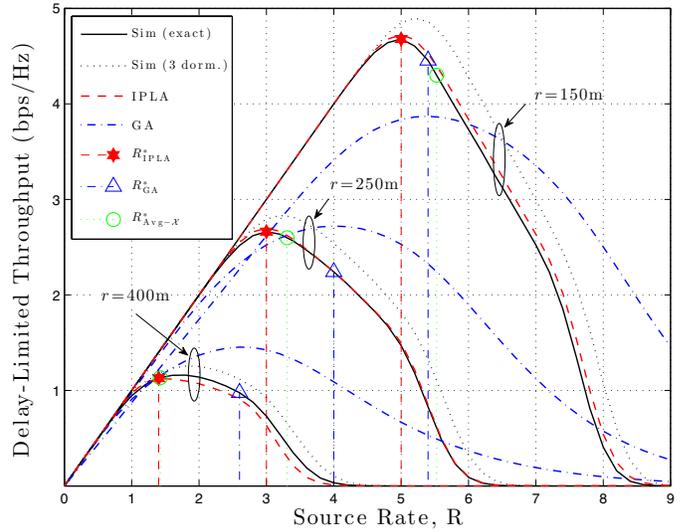}
    \caption{Effect of User Distance on Rate Selection ($K=6$, $\theta=\pi/2$, $\alpha=3$, $|w_{u_0}^{(0)}|^2=1$, $N_{max}=4$) }
    \label{fig:effect_dist}
\end{figure}

In this subsection, we investigate the effects of user distance and path-loss exponent on the rate selection according to the proposed and conventional link adaptation schemes.
Fig.~\ref{fig:effect_dist} shows the DLT for varying source rate $R$ in three different user distance values. Basically, the DLT has a shape of quasi-concave function and a single optimal point with respect to the source rate. As the distance decreases, a higher DLT is achieved since the desired signal power increases while a closer distance to the BS fundamentally yields smaller interference from other-cell BSs. In the figure, the solid lines denote the exact simulation results with perfectly known individual ICIs. Additionally, the dotted lines denote the simulation results with perfectly known three dominant ICIs, which neglect the other three ICIs. Compared to both simulation results, neglecting non-dominant ICIs yields over-estimated DLTs due to the reduced interference even if it shows similar shapes of curves. On the contrary, the DLT analytically derived by the proposed IPLA has a high similarity with  one by the exact simulation for all distance values. After all, the optimal source rate determined by the proposed IPLA is approximately identical to the actual optimal source rate on the exact simulation curves, regardless of the user distances. 
However, optimal source rates based on the two conventional link adaptation schemes, GA and average interference schemes, exhibit significant differences from the actual optimal value on the exact simulation curves. The gap between the optimal source rate by the proposed IPLA and one by the conventional GA increases as the user distance increases (i.e., as ICI increases), whereas the gap between the optimal source rate by the proposed IPLA and one by the conventional average interference scheme increases as the user distance decreases. Therefore, we can conclude that the conventional GA is relatively good for near-BS users while the conventional average interference scheme is good for edge users, in the link adaptation perspective.

\begin{figure}
    \centering
    \includegraphics[width=\columnwidth]{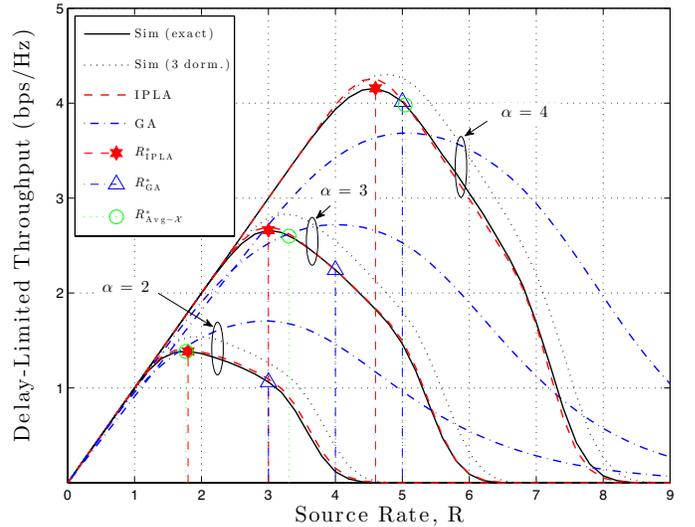}
    \caption{Effect of Path-Loss Exponent on Rate Adaptation ($K=6$, $r=250$ m, $\theta=\pi/2$, $|w_{u_0}^{(0)}|^2=1$, $N_{max}=4$)}
    \label{fig:effect_plexp}
\end{figure}

Fig.~\ref{fig:effect_plexp} shows the DLT for varying source rate $R$ in three different path-loss exponent values. As the path-loss exponent value increases, a higher DLT is achieved since the interference is reduced with increasing the path-loss exponent value. The optimal source rate by the proposed IPLA also agrees well with the actual optimal value on the exact simulation curve, while those by the conventional schemes show significant differences. The basic trends of the differences are similar to those in Fig.~\ref{fig:effect_dist}. Consequently, for the conventional GA, the more interference exists, the larger difference occurs in the optimal source rate, while for the conventional average interference scheme, the less interference exists, the larger difference occurs in the optimal source rate. Fundamentally, both the conventional schemes exhibit significant differences with respect to the optimal source rate for large SINR values (i.e., $\alpha=4$), which correspond to the weak interference situation. However, in a scheduling-based multi-user system, a user with a large SINR value has more opportunities to be selected as the best user. Therefore, it is expected that the proposed IPLA-based cross-layer policy can obtain a significant throughput gain in the viewpoint of both link adaptation and user scheduling, compared to the conventional policies.

\subsection{System-Level Performance Evaluation: Cell Throughput and Fairness}

\begin{figure}
    \centering
    \subfigure[]{
        \includegraphics[width=\columnwidth]{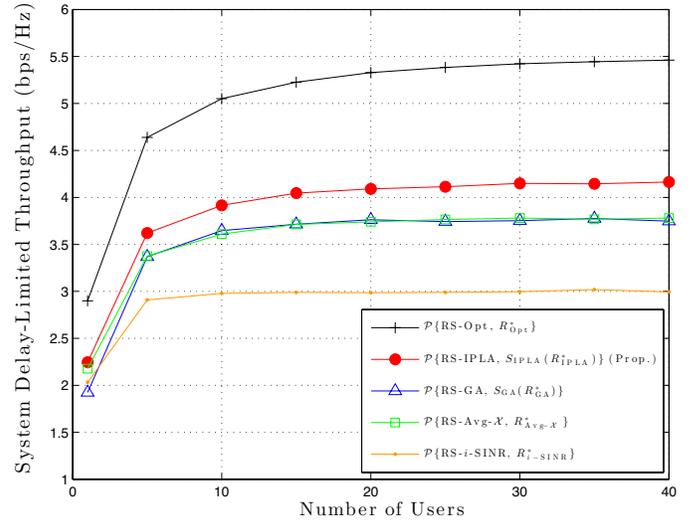}}
    \subfigure[]{
        \includegraphics[width=\columnwidth]{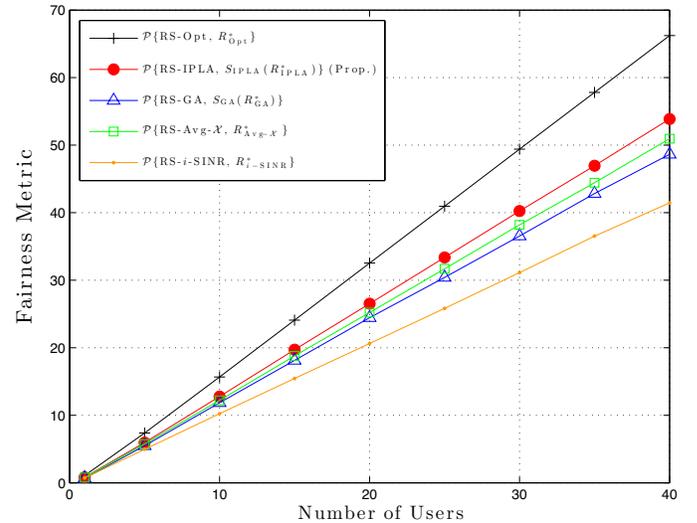}}
    \caption{System Performance in an Asymmetric User Distribution Scenario (a) System DLT vs. Number of Users (b) Fairness Metric vs. Number of Users ($K=6$, $r\in [150,200,250,300,400]$m, $\theta=$Uniform$[-\pi,\pi]$, $\alpha=3$, $N_{max}=4$)}
    \label{fig:asym_user_dist}
\end{figure}

Fig.~\ref{fig:asym_user_dist} shows the system performance of various cross-layer policies for varying the number of users in the home-cell. Specifically, Fig.~\ref{fig:asym_user_dist}~(a) shows the system throughput of the proposed, conventional, and reference policies. The genie-aided policy, $\mathcal{P}$\{RS-Opt, $R_{\mathrm{Opt}}^*$\}, provides an upper bound of the system throughput even if it is unrealistic. The instantaneous SINR policy, $\mathcal{P}$\{RS-$i$-SINR, $R_{i\text{-SINR}}^*$\}, achieves the worst system throughput due to rather inaccurate estimation of the ICI term caused by the channel feedback delay. The proposed IPLA-based policy, $\mathcal{P}$\{RS-IPLA, $S_{\mathrm{IPLA}}(R_{\mathrm{IPLA}^*})$\}, always outperforms the conventional policies, $\mathcal{P}$\{RS-GA, $S_{\mathrm{GA}}(R_{\mathrm{GA}}^*)$\} and $\mathcal{P}$\{RS-Avg-$\mathcal{X}$, $R_{\text{Avg-}\mathcal{X}}^*$\}, in the entire range of the number of users, while both the GA and average interference policies achieve almost identical system throughput. Note that although the GA-based policy exploits statistics of the ICI term, it achieves similar system throughput to that of the average interference policy which just utilizes an average value of the ICI term. Moreover, it achieves rather smaller system throughput than that of the proposed IPLA-based policy which also exploits equivalent average statistics of the ICI term.

Fig.~\ref{fig:asym_user_dist}~(b) shows the fairness metric performance of various cross-layer policies for varying the number of users in the home-cell. 
We consider the \emph{fairness metric} in \cite{s_BL99_Tse,p_TIT02_Viswanath,p_TWC05_Zheng,p_TVT13_Kim} defined as $\mathcal{FM}(T_1,\ldots,T_N)=\sum_{u=1}^{N}\log(T_u)$ where $T_u$ denotes the achieved throughput of the $u$-th user and $N$ is the number of users in the system. As investigated in the previous work, the fairness metric offers a performance measure considering both system throughput and user fairness together. Under the PF scheduling algorithm with averaging time scale $t_c=\infty$, the fairness metric is maximized almost surely among the class of all schedulers \cite{p_TIT02_Viswanath}. As shown in Fig.~\ref{fig:asym_user_dist}~(b), the proposed IPLA-based policy, $\mathcal{P}$\{RS-IPLA, $S_{\mathrm{IPLA}}(R_{\mathrm{IPLA}}^*)$\}, also outperforms the other three policies except for the genie-aided policy for all the number of users. The average interference policy, $\mathcal{P}$\{RS-Avg-$\mathcal{X}$, $R_{inst}^*$\}, rather outperforms the GA-based policy, $\mathcal{P}$\{RS-GA, $S_{\mathrm{GA}}(R_{\mathrm{GA}}^*)$\}, in terms of the fairness metric, even if the GA-based policy exploits more information for ICI term than the average interference policy. 
It comes from the fact that the GA on the aggregated ICI is inaccurate when there exist some dominant ICIs which are general in OFDM-based cellular networks, although the GA on the aggregated ICI is well-approximated for a sufficiently large number of independent and identically distributed interferers. In contrast, the proposed IPLA is highly accurate in this environment. Accordingly, the proposed IPLA-based cross-layer policy is able to be the most efficient in the OFDM-based cellular networks where there exist some dominant ICIs.

\section{Conclusion} \label{SEC:ICI:Conclusions}

In this paper, we investigated a joint link adaptation and user scheduling in multi-user and multi-cell environments, considering HARQ techniques. Based on the proposed mathematical approximation method for the ICI signals the optimal transmission rate selection algorithm in terms of the expected throughput is proposed. As for multi-user environments, a novel and effective cross-layer framework combining the link adaptation and user scheduling is also proposed. Through extensive link-/system-level simulations, it is shown that the proposed cross-layer policy significantly outperforms the conventional policies in terms of both cell throughput and user fairness. 
With consideration of both HARQ and ICI, we have tried to investigate more general and practical communication scenarios including multi-user MIMO, receiver beamforming at users with multiple receive antennas, and HARQ technique with incremental redundancy, but they were not mathematically tractable unfortunately.
Therefore, we leave these issues for future work.

\balance

\bibliographystyle{./style/IEEEtran_v111}
\bibliography{./style/IEEEabrv,./style/RefAbrv,Reference}

\end{document}